\documentclass[epsf]{aa}
\usepackage{graphicx}
\begin{document}
   \title{The FIR/Radio correlation of high redshift galaxies \\ 
          in the region of the HDF-N}


   \author{M.A. Garrett 
          \inst{}
          }

   \offprints{M.A. Garrett}

   \institute{Joint Institute for VLBI in Europe, Postbus 2, 7990~AA
     Dwingeloo, NL\\
   \email{garrett@jive.nl}}

   \date{}

   \abstract{The correlation between the far-infrared (FIR) and radio
     emission is well established for nearby star forming galaxies.
     Many applications, in particular the radio-to-submm spectral index
     redshift indicator, tacitly assume that the relation holds well
     beyond our local neighbourhood, to systems located at cosmological
     distances. In order to test this assumption I have constructed a
     sample of 20 HDF-N galaxies, all with measured spectroscopic
     redshifts, and all detected by {\it both} ISO and the WSRT at
     15~micron and 1.4~GHz respectively.  The galaxies span a wide
     range of redshift with a median value of $z \sim 0.7$. The ISO 15
     micron data were k-corrected and extrapolated to the FIR (60 and
     100 micron) by assuming a starburst (M82) spectral energy
     distribution (SED) for the entire sample. An initial analysis of
     the data suggests that the correlation between the FIR and the
     radio emission continues to apply at high redshift with no obvious
     indication that it fails to apply beyond $z \sim 1.3$.  The sample
     is ``contaminated'' by at least 1 distant ($z=4.4$), radio-loud
     AGN, VLA~J123642+621331. This source has recently been detected by
     the first deep field VLBI observations of the HDF-N and is clearly
     identified as an outlier in the FIR/radio correlation.
      }

   \maketitle
%

\section{Introduction}

The correlation between the far-infrared (FIR) and radio emission is
one of the tightest and most universal correlations known among the
global properties of {\it local} star forming and starburst galaxies
(see van der Kruit 1973, Helou \& Bicay 1993 and references therein).
Entirely unexpected, yet extending across five orders of magnitude in
luminosity, the physical explanation for the tightness of the relation
is that the non-thermal radio emission and the thermal FIR emission are
both related to processes governed by massive star formation.
Established via FIR and radio observations of {\it nearby} galaxies,
the FIR/radio correlation forms the basis of several key applications.
In particular, the technique of using the radio-to-submm spectral index
as a redshift indicator (Carilli \& Yun, 2000) and the use of unbiased
radio observations to estimate the global star formation history of the
Universe (Haarsma et al.  2000). These applications are beginning to
overhaul current ideas of star formation and galaxy formation in the
early Universe. In particular, deep SCUBA sub-mm observations of the
high redshift Universe (e.g. Hughes et al. 1998, Barger et al. 1999)
reveal a dominant population of dusty, optically faint (early type)
galaxies that are inferred (via the radio-to-submm spectral index
redshift indicator) to lie at cosmological distances ($z>2$). These
results are in sharp contrast to optical and ultraviolet studies which
argue for a strong peak in the global star formation rate at around
$z\sim 1$ (e.g. Madau et al. 1996).

\begin{table*}[ht] 
\centering
\caption{Details of the WSRT/ISO Source Sample}
\begin{tabular}{|lllllll} 
\label{table1} 
WSRT ID & RA ($+ 12$ hr) & DEC ($+ 62^{\circ}$) & S$_{1.4 {\rm GHz}}$ & ISO ID & 
S$_{15\mu}$ & $z$ \\ 
   & (m,s) & ($^{\prime}$ $^{\prime\prime}$)& $\mu$Jy &  &
$\mu$Jy &   \\ 
\hline 
3631+1108 & 36 31.59 & 11 08.20 & $82\pm26$ & HDF\_PM3\_1 & 355$^{+40}_{-60}$ & 1.013 \\ 
3634+1239 & 36 34.44 & 12 39.63 & $185\pm22$ & HDF\_PM3\_3 & 363$^{+79}_{-38}$ & 1.219 \\ 
3634+1214 & 36 34.46 & 12 14.26 & $254\pm23$ & HDF\_PM3\_2 & 448$^{+68}_{-59}$ & 0.458 \\  
3635+1427 & 36 35.09 & 14 27.16 & $124\pm26$ & HDF\_PM3\_5 & 441$^{+43}_{-82}$ & 0.297 \\  
3636+1132 & 36 36.78 & 11 32.07 & $77\pm25$ & HDF\_PM3\_7 & 300$^{+62}_{-67}$ & 0.078 \\ 
3637+1151 & 36 37.32 & 11 51.89 & $151\pm49$ & HDF\_PM3\_10 & 212$^{+58}_{-55}$ & 0.842 \\ 
3637+1117 & 36 37.53 & 11 17.19 & $62\pm26$ & HDF\_PM3\_9  & 212$^{+58}_{-55}$ & 1.018 \\ 
3644+1247 & 36 44.19 & 12 47.31 & $120\pm30$ & HDF\_PM3\_17 & 282$^{+60}_{-64}$ & 0.557 \\ 
3646+1405 & 36 46.02 & 14 05.63 & $187\pm22$ & HDF\_PM3\_20 & 107$^{+95}_{-20}$ & 0.962 \\ 
3646+1528 & 36 46.19 & 15 28.39 & $81\pm23$ & HDF\_PM3\_21 & 418$^{+91}_{-94}$ & 0.851 \\ 
3646+1447 & 36 46.20 & 14 47.56 & $257\pm28$ & HDF\_PM3\_23 & 144$^{+72}_{-47}$ & 0.558 \\
3647+1427 & 36 47.84 & 14 27.02 & $116\pm41$ & HDF\_PM3\_24 & 307$^{+62}_{-67}$ & 0.139 \\ 
3649+1314 & 36 49.58 & 13 14.12 & $68\pm19$ & HDF\_PM3\_27 & 320$^{+39}_{-62}$ & 0.475 \\
3650+1034 & 36 50.98 & 10 34.63 & $72\pm22$ & HDF\_PM3\_28 & 341$^{+40}_{-65}$ & 0.410 \\
3651+1357 & 36 51.36 & 13 57.04 & $82\pm26$ & HDF\_PM3\_30 & 151$^{+74}_{-68}$ & 0.557 \\ 
3651+1222 & 36 51.81 & 12 22.37 & $57\pm16$ & HDF\_PM3\_29 &  48$^{+32}_{-9}$ & 0.401 \\ 
3653+1137 & 36 53.05 & 11 37.67 & $73\pm21$ & HDF\_PM3\_32 & 180$^{+60}_{-43}$ & 1.275 \\ 
3702+1401 & 37 02.40 & 14 01.24 & $55\pm23$ & HDF\_PM3\_44 & 144$^{+73}_{-47}$ & 1.243 \\
3641+1546 & 36 41.90 & 15 46.87 & $98\pm25$ & HDF\_PS3\_10 & 459$^{+46}_{-86}$ & 0.857 \\
3641+1331 & 36 41.90 & 13 31.96 & $489\pm22$ & HDF\_PS3\_6e & 23$^{+10}_{-12}$ & 4.424 \\ 
\hline
\end{tabular} 
\end{table*} 

These sub-mm and radio based claims for substantial global star
formation rates at high redshift depend upon a crucial assumption: {\it
  viz.}  that the FIR/radio correlation is entirely independent of
redshift.  In fact there are many reasons why the correlation might
well fail at non-local redshifts. In the radio domain these include
primarily (Condon 1992, Lisenfeld, Volk \& Xu 1996), the quenching of
the radio emission due to inverse Compton (IC) losses via the CMB
(scaling as $(1+z)^{4}$). In addition, the trend for higher redshift
systems to be more luminous (a selection effect related to current
sensitivity limits) may also lead to global changes in the properties
of the sources actually detected (as compared to local, less luminous
star forming galaxies that form the basis of the locally derived
relation).  An overall change in the SED of these systems might well be
expected, and the differing time scales associated with the rise in the
FIR and radio emission may be significant, particularly for vigorous
starburst systems. Specific effects regarding the level of radio
emission might occur via (i) evolving magnetic field strengths (Dunne
et al. 2000), (ii) varying levels of free-free absorption (Rengarajan
\& Takeuchi 2001), (iii) further IC losses associated with the intense
but {\it local} radiation environment and (iv) the possibility of
significant ``contamination'' of high redshift star forming systems
with co-existing, low-luminosity, ``radio loud'' AGN.

Until recently a study of the FIR/radio correlation at non-local
redshifts has been difficult since it requires extremely deep and
complementary observations in both the radio and far/mid-IR wavebands,
together with spectroscopic redshifts of relatively faint sources. Like
many other areas of astrophysics, the situation has recently been
transformed by the wealth of publicly available data generated by
deep multi-wavelength studies of the Hubble Deep Field (HDF). In this
paper I investigate the nature of the FIR/radio correlation at moderate
redshifts (up to $z\sim 1.3)$ assuming throughout the currently
``preferred'' cosmological model ($\Omega_{{\rm m}}=0.3$,
$\Omega_{\Lambda}=0.7$, $H_{{\rm 0}}=70$~km/sec/Mpc).

   \begin{figure*}[ht]
   \centering
   \includegraphics[scale=0.65,angle=0]{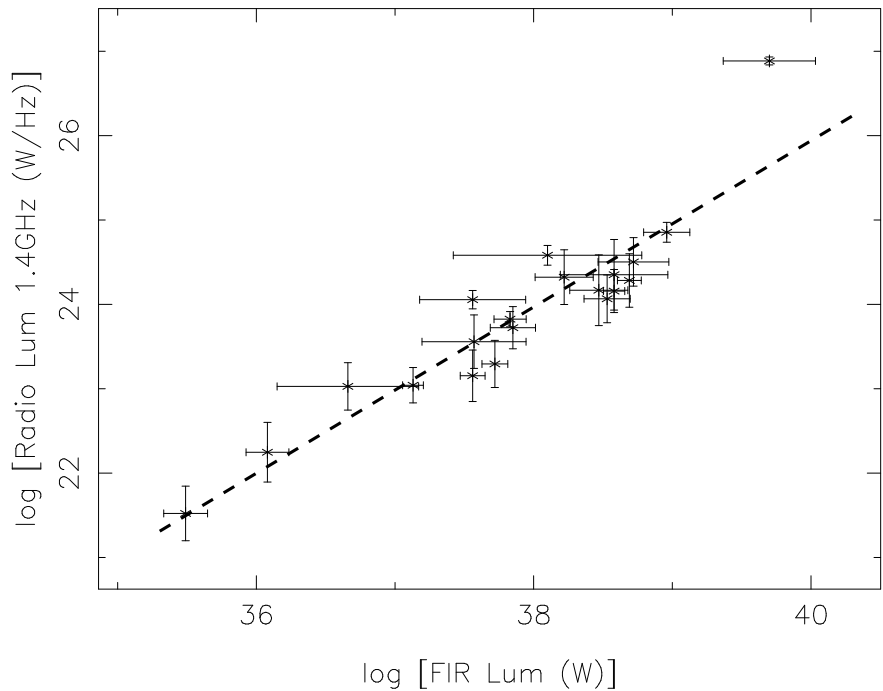}
   \caption{
A logarithmic plot of the Radio vs FIR luminosity of the
  sample. A median fit to the data (dashed line) is also presented - 
  the FIR/radio correlation is clearly seen to apply out to
  $z\sim1.3$. The distant ($z=4.4$) radio-loud AGN 
  shows up as an outlier at the far RHS of the plot.
}
              \label{Fig1}
    \end{figure*}

\section{The ISO 15 micron and WSRT 1.4~GHz HDF-N Sample} 

A sample of 20 HDF-N sources was established from the following simple
criteria: (i) they are detected by {\it both} the WSRT at 1.4~GHz
(Garrett et al. 2000) and ISO at 15 micron (Aussel et al. 1999), and
(ii) each WSRT/ISO source is independently identified with the same
optical candidate with a measured redshift (as determined by Cohen et
al. 2001). Details of the sample is presented in Table 1.  The WSRT
sources are all detected with a SNR $> 5\sigma$. The ISO 15 micron
sources are mostly drawn from the PRETI main source list (Aussel et al.
1999) but a few of the most significant (SNR $> 4\sigma$) detections
presented in the PRETI supplementary source list are also included. No
attempt was made to remove possible AGN candidates from the sample.

One limitation of this study is that the deep ISO measurements were
made in the mid-infrared and it was therefore necessary to extrapolate
the 15 micron measurements to 60 and 100 micron. This was achieved by
constructing a SED based on the starburst galaxy M82 (this source has
been extensively studied at all wave-bands including those that form the
basis of our SED - the radio, mm, sub-mm, far, mid and near infrared
bands). A k-correction (dependent on source redshift and our assumed
SED) was also applied to both the mid-infrared and radio data. Note
that for non-local redshifts the steepness of the Rayleigh-Jeans tail
in the FIR makes the k-correction absolutely essential, without this
the {\it observed} FIR-Radio correlation completely disappears (it is
this same property that, in principle, makes the sub-mm/radio spectral
index such a powerful redshift estimator). I adopt the FIR/radio ratio
as quantified in the $q$ parameter defined by Condon (1992):
$q={\rm log}(\frac{S_{100}+2.6S_{60}}{3S_{1.4}})$

\section{Results} 

Figure~1 shows a logarithmic plot of the FIR and radio luminosities for
the HDF-N sample.  A median-fit to the data indicates the linearity of
the relation (the slope of the fit is 1.00 with a correlation
coefficient of 0.93). This striking result strongly suggests that the
FIR/radio correlation continues to apply at non-local, moderate
redshifts.  Note that the luminosities probed by our faint (but
distant) sample extends the range (upwards) by two orders of magnitude
compared to previous, ``local'' studies. The highest luminosity sources
in Figure~1 also tend to be more distant.

Figure~2 shows the value of $q$ plotted as a function of redshift.
There is a clear clustering of sources around $q\sim 2.0$ (excluding
the radio-loud outliers). Given that the dominant error is the process
of extrapolating the mid-infrared fluxes to the far-infrared via our
assumed SED, this is remarkably similar to the value measured by Condon
(1992) for local galaxies - $q\sim 2.3$. The absolute value of $q$
quoted here should not be over interpreted: for example repeating the
analysis with an Arp 220 SED increases the value of $q$ to 2.3. 

There is one obvious outlier observed in both Figure~1 and 2.  This
outlier (seen to the extreme right of both figures) corresponds to a
$z=4.4$ source (VLA~J123642+621331), and is widely believed to be an
optically faint, dust obscured star forming galaxy, which also harbours
an AGN (Waddington et al. 1999, Garrett et al. 2001, Brandt et al.
2001). The fact that it deviates from the FIR/radio correlation is
entirely to be expected, and suggests that just as the locally derived
correlation can distinguish between sources powered by AGN and star
formation processes, the extended correlation can very likely do the
same thing for moderate redshift sources.

   \begin{figure*}
   \centering
   \includegraphics[scale=0.65,angle=0]{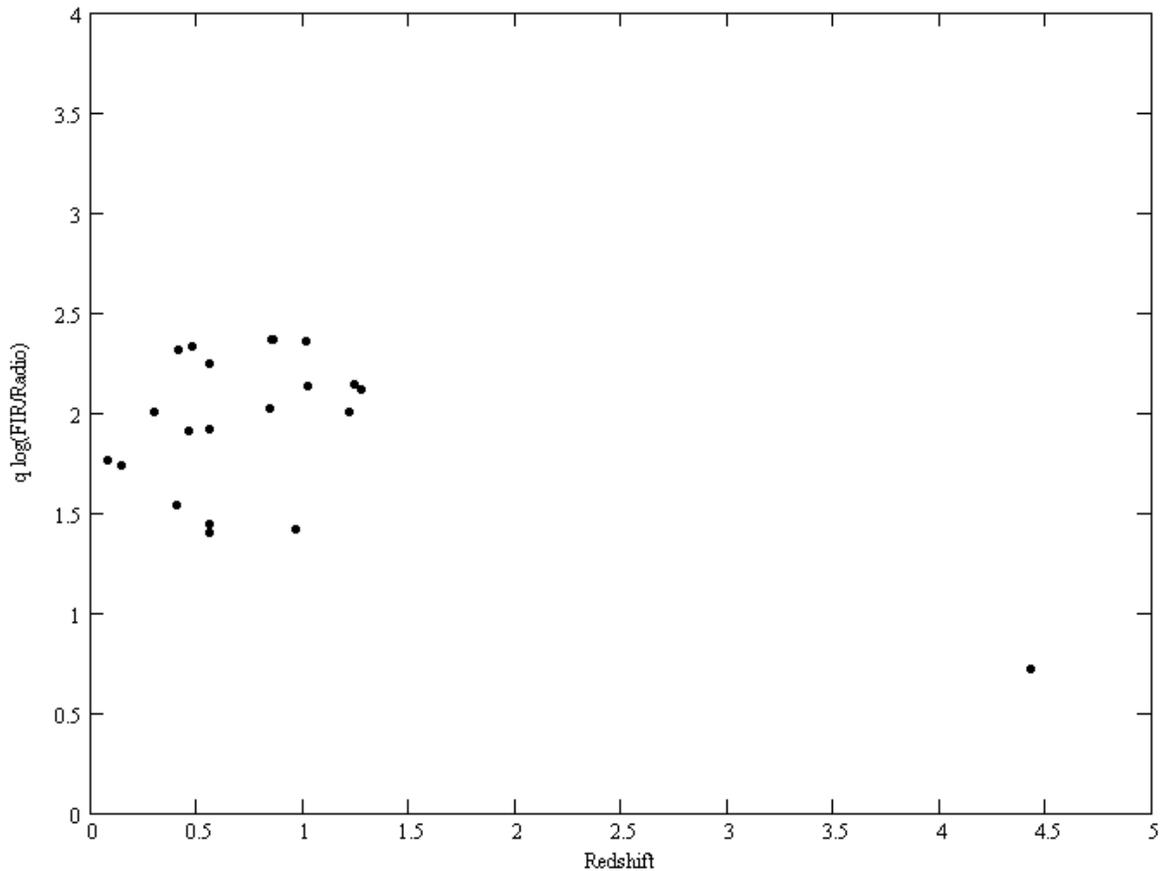}
   \caption{
     The ratio, $q$ of the FIR/Radio emission plotted against redshift
     for each galaxy in the sample.  The distant ($z=4.4$) radio-loud
     AGN (extreme RHS) clearly shows up as an outlier in this plot.}
              \label{Fig2}
    \end{figure*}

\section{Conclusions} 

The FIR/radio correlation appears to apply to the vast majority of the
20 HDF-N galaxies in our sample. There is no obvious evidence to
suggest that the relation does not apply equally well, beyond $z\sim
1$. This result lends support to the idea (Carilli \& Yun 2000) that
the radio-to-submm spectral index is a useful redshift indicator for
the optically faint, sub-mm and radio source population. In addition,
it re-emphasises the crucial role that deep radio observations play as
an unbiased estimator in studies of the star-formation history of the
(obscured) early Universe.

Our analysis extends the FIR/radio correlation (in terms of source
luminosity) by up to 2 orders of magnitude (as compared to local
studies). In addition to the puzzle of just why the FIR/radio
correlation is so tight in the local Unverse, it seems we must also
consider why it continues to apply equally well to these more luminous
and more distant systems. A more extensive investigation of the
FIR/radio correlation at high redshift awaits much deeper SIRTF and
ALMA infrared and sub-mm observations. Complimentary radio data will be
essential and these depend critically on upgraded instruments such as
those currently proposed ($e$-MERLIN and EVLA). On longer time scales,
next generation radio instruments, in particular the Square Km Array
(SKA), will {\it completely} revolutionise our view of galaxy formation
in the early Universe.  SKA will detect even relatively feeble/normal
($\sim$~few~$M_{\odot}$/yr) star forming galaxies out to any redshift
that they might reasonably be expected to exist.

\begin{acknowledgements}
  I would like to thank the referee Min S. Yun for useful comments and
  suggestions. The Westerbork Synthesis Radio Telescope is operated by
  the ASTRON (Netherlands Foundation for Research in Astronomy) with
  support from the Netherlands Foundation for Scientific Research, NWO.
  The European VLBI Network is a joint facility of European and Chinese
  radio astronomy institutes funded by their national research
  councils.

\end{acknowledgements}

\end{document}